%% file: mnras_template.tex
\documentclass[fleqn,usenatbib]{mnras}

\usepackage{newtxtext,newtxmath}

\usepackage[T1]{fontenc}

\DeclareRobustCommand{\VAN}[3]{#2}
\let\VANthebibliography\thebibliography
\def\thebibliography{\DeclareRobustCommand{\VAN}[3]{##3}\VANthebibliography}

\usepackage{graphicx}	
\usepackage{amsmath}	

\usepackage[dvipnames]{xcolor}

\title[SNIa systematics: $w_0$-$w_a$]{The axis of systematic bias in SN~Ia cosmology and implications for DESI 2024 results}
\author[]{
Suhail Dhawan,$^{1}$\thanks{E-mail:sd919@cam.ac.uk, suhail.dhawan@gmail.com}
Brodie Popovic,$^{2}$
Ariel Goobar$^{3}$
\\
$^{1}$Institute of Astronomy and Kavli Institute for Cosmology, University of Cambridge, Madingley Road, Cambridge CB3 0HA, UK\\
$^{2}$Universite Claude Bernard Lyon 1, CNRS, IP2I Lyon / IN2P3, IMR 5822, F-69622 Villeurbanne, France\\
$^{3}$The Oskar Klein Centre for Cosmoparticle Physics, Department of Physics, Stockholm University, SE-10691 Stockholm, Sweden
}

\date{Accepted XXX. Received YYY; in original form ZZZ}

\pubyear{\the\year{}}

\begin{document}
\label{firstpage}
\pagerange{\pageref{firstpage}--\pageref{lastpage}}
\maketitle

\begin{abstract}
Relative distances between a high-redshift sample of Type Ia supernovae (SNe~Ia), anchored to a low-redshift sample, have been instrumental in drawing insights on the nature of the dark energy driving the accelerated expansion of the universe. A combination (hereafter called SBC) of the SNe~Ia  with baryon acoustic oscillations (BAO) from the Dark Energy Spectroscopic Instrument (DESI) and the cosmic microwave background (CMB) recently indicated deviations from the standard interpretation of dark energy as a cosmological constant. In this paper, we analyse various systematic uncertainties in the distance measurement of SNe~Ia and their impact on the inferred dark energy properties in the canonical Chevallier-Polarski-Linder (CPL) model. We model systematic effects like photometric calibration, progenitor and dust evolution, and uncertainty in the galactic extinction law. We find that all the dominant systematic errors shift the dark energy inference towards the DESI 2024 results from an underlying $\Lambda$CDM cosmology.  A small change in the calibration, and change in the Milky Way dust, can give rise to systematic-driven shifts on $w_0$-$w_a$ constraints, comparable to the deviation reported from the DESI 2024 results. We forecast that the systematic uncertainties can shift the inference of $w_0-w_a$ by a few times the error ellipse for future low- and high-$z$ SN~Ia compilations and hence, it is critical to circumvent them to robustly test for deviations from $\Lambda$. A slider and visualisation tool for quantifying the impact of systematic effects on the fitted cosmological parameters is publicly  available at: \hyperlink{slider}{https://github.com/sdhawan21/DEslider.git}
\end{abstract}




\section{Introduction}
Over the past quarter-century, a wide variety of cosmological measurements have lent further support for what has become the standard cosmological model \citep[termed as the $\Lambda$CDM model, e.g.,][]{Peebles2024}: a spatially flat universe with an energy density today composed of about 5\% baryonic matter, 25\% cold dark matter (CDM), and 70\% dark energy. Understanding the  physical origin of this `dark energy' is a fundamental question in modern cosmology. While the leading explanation for dark energy is the cosmological constant ($\Lambda$), theoretical problems with this explanation \citep[e.g.,][]{Weinberg2000} have motivated a more detailed exploration of possible origins via precision measurements from independent observational probes \citep[e.g.,][]{Lovick2023,Calderon2024,Camilleri2024}. 

The magnitude-redshift relation of  Type Ia supernovae (SN~Ia) provided the first direct observational evidence for accelerating expansion \citep{Riess1998,Perlmutter1999}, see \citet{2011ARNPS..61..251G} for a review. SNe~Ia\footnote{In this paper SN~Ia refers to constraints from the magnitude-redshift relation} are  a pivotal probe in precisely measuring the properties of dark energy \citep{Brout2022_cosmo,Rubin2023,DES2024}, e.g. its present-day equation of state ($w_0$) and time-dependence ($w_a$)  - hereafter termed as $w_0$ - $w_a$ - with some compilations being a culmination of decades of different surveys \citep{Kessler2009,Sullivan2011,2014A&A...568A..22B}. Combining the SNe~Ia with the temperature and polarisation fluctuations of the cosmic microwave background \citep[CMB;][]{Planck2020} and late-universe measurement of the acoustic peak in the clustering of galaxies (baryon acoustic oscillations \citep[(BAO);][]{DESI-VI} has been the most precise route to constrain the properties of dark energy. There are several next-generation experiments, termed as stage-IV based on the Dark Energy Task Force convention \citep{Albrecht2006}, that are planned to significant improve the empirical constraints on dark energy \citep{Laureijs2011,Ivezic2019,Scolnic2019} in the near-future. Therefore, it is timely to interpret results from current experiments and quantify how present-day systematic uncertainties can impact dark energy inference today, and in the near-future.
\vspace{-0.03cm}

The most precise current constraints on $w_0 - w_a$ are from a combination of the latest SNe~Ia compilations \citep{Brout2022_cosmo,Rubin2023,DES2024} with BAO from the Dark Energy Spectroscopic Instrument \citep[DESI;][]{DESI-VI} and the CMB constraints from {\emph Planck}. The results suggest deviations from the fiducial $\Lambda$CDM cosmology, i.e. ($w_0$, $w_a$) of (-1, 0) between 2.6 - 3.9 $\sigma$ significance depending on the SN~Ia sample used. The inference of the $w_0$ and $w_a$ from SNe~Ia relies on the relative distance measurements between high-$z$ ($z > 0.1$) SNe~Ia and their low-$z$ ($z \leq 0.1$) counterparts. Therefore, to control the systematic uncertainty budget, it is crucial to have a robust understanding of any non-uniformities in the SN~Ia population across the full redshift. This includes a precise understanding of both instrumental effects, such as cross-telescope calibration, and astrophysical effects, such as circum- and inter-stellar dust. In this paper, we explore the different sources of systematics and how they can impact $w_0 - w_a$ constraints. This also sets the target precision for robustly inferred $w_0 - w_a$ constraints from future surveys.

Historically, one of the key sources of systematic uncertainties has been the calibration of low-$z$ SNe~Ia, which have typically been observed with several heterogeneous systems, rendering a complete  model of their selection functions prohibitively difficult. This is particularly important since SNe~Ia are the only precision cosmology probe in the $z \lesssim 0.1$ range \citep[see also][]{Efstathiou2024b}. Today, the problem of a large, well-calibrated low-$z$ anchor sample looks to be overcome in the current era of wide-field surveys with the Zwicky Transient Facility \citep[ZTF;][]{Graham2019,Dhawan2022,Rigault2024}. ZTF has already, in its first phase of operations, characterised a sample of $\sim 2500$ cosmological-grade SNe~Ia at $z \leq 0.1$. Combining this sample with the state-of-the-art high-$z$ surveys \cite{Sanchez2024,Vincenzi2024,DES2024} will allow us to constrain the dark energy equation-of-state and its potential evolution with cosmic time: $w_0$-$w_a$. In this paper, we quantify major sources of systematics and their impact on $w_0$-$w_a$. These systematic uncertainties will determine the degree to which future surveys and analyses with SNe~Ia will be able to achieve their cosmology programs, and surpass the key metrics laid out by the Dark Energy Task Force convention and discern further, and more complex, cosmological parameters with future high-$z$ samples \citep{Hounsell2018,lsst_srd}. 


To test these systematic uncertainties, we follow the spirit of the work by \citet{Nordin2008}, who perturbed simulated distances from a combined low-$z$ and high-$z$ sample with the sample size and uncertainty budget similar to current samples. 
In this paper, we use the covariance from the current most-precise SN~Ia compilation \citep{Brout2022_cosmo}, focusing on instrumental (e.g. calibration, see \citet{Brout2022}) and astrophysical systematics, e.g., differences in the dust distribution with redshift,  \citep[][]{BS21,Popovic2021}, intrinsic scatter \citep[e.g.,][]{Popovic2023}, Milky Way extinction laws, intergalactic dust \citep{Goobar2018} and possible progenitor evolution. We consider the impact of these different systematic groupings and present the expected constraints.

The paper is outlined as such: Section \ref{sec:data} provides a brief outline of the data; Section \ref{sec:method} describes the inference of dark energy parameters, followed by Section \ref{sec:sys-group}, which lists
all the systematic effects with their impact on dark energy inference summarised in Section~\ref{sec:impact_de}. We summarise our findings in Section~\ref{sec:discussion_conclusions}.

\section{Data}\label{sec:data}
We make use of the redshifts and covariance matrix from the Pantheon+ cosmology analysis \citep{Brout2022_cosmo, Scolnic2022}. Pantheon+ constitutes $\sim 1500$ unique SNe Ia, spanning a range from $0.01 < z < 2.26$, and is to-date the most accurate measurement of the dark energy equation-of-state \textit{w} to-date. The Pantheon+ data is publicly available \footnote{\hyperlink{panth}{https://github.com/PantheonPlusSH0ES/DataRelease}}. For our analysis, we simulate distances based off of an assumed cosmology and redshift (see Section~\ref{sec:method}), in tandem with the covariance matrix for estimating our error budget.
\vspace{-0.2cm}
\subsection{External Data}
. The most precise constraints on $w_0-w_a$ come from the combination of the SN~Ia magnitude-redshift relation with complementary probes, e.g. BAO and CMB. Since the focus of this work is the residual systematic uncertainties in the SNe~Ia, we refer to the complementary probes as ``external data". This is mostly due to independent constraints on the present-day matter density, and the difference in the angle of the $w_0-w_a$ degeneracy from BAO compared to SNe~Ia. For the CMB, we use the compressed likelihood with the same input cosmology as for the SNe~Ia \citep{Chen2019}. The data vector for includes the CMB shift, $R$, position of the first acoustic peak in the power spectrum, $l_{\rm A}$ and the baryon density at present day, $\Omega_{\rm b} h^2$ comprise the data vector. The expression for the CMB shift and the position of the first acoustic peak are given by

\begin{equation}
R = \sqrt{\Omega_\mathrm{M} H_0^2} d_{\rm A}(z_*)/c,
\label{eq:cmb_shift}
\end{equation}
and
\begin{equation}
l_{\rm A} = \pi \frac{d_{\rm A}(z_*)}{r_s(z_*)}
\label{eq:cmb_la}
\end{equation}
where $r_s(z)$ is the sound horizon at redshift, $z$ \citep[see][for details]{Chen2019}. We emphasize that since we generate the mock observables for a $w_0$-$w_a$ cosmology, the use of the compressed likelihood is valid \citep[as discussed in detail in][]{Planck2015_DE}. We compute the above synthetic observables for the CMB compressed likelihood using a true cosmology of ($\Omega_M$, $w_0$, $w_a$) of (0.3, -1, 0) as well as an $H_0$ of 70 km/s/Mpc, however, since we also marginalise the absolute luminosity of SNe~Ia, this does not impact our results. We tested this by changing $H_0$ and re-running the inference and did not find a difference in the inferred $\Omega_M, w_0, w_a$ 
BAO analyses use a few different distance measurements depending on the dataset. These are $D_{\rm A}$, $D_{\rm C}$, and $D_{\rm V}$, i.e. the angular diameter, comoving and volume distances, which we collectively refer to as $D_x$. While we simulate the distances at each redshift for our input cosmology, we use the distance errors from the combined DESI+SDSS BAO compilation \citep{DESI-VI}. The redshifts and errors are summarised in Table~\ref{tab:bao_data}. We use the priors on the cosmological parameters from \citet{DESI-VI}. For the combined probes, which is the focus of this paper, the constraints are very stringent and not dependent on the prior range.
\input{tables/bao_test}

\section{Methodology}
\label{sec:method}
In this section we describe the  inference methodology using the SN~Ia redshifts, simulated distances and covariance matrix to obtain the cosmological parameters.  The distance modulus predicted by a homogeneous and isotropic, flat Friedman-Robertson-Walker (FRW) universe is given by

\begin{equation}
\mu(z; \theta) = 5\, \mathrm{log_{10}} \left( \frac{D_{\rm L}}{10\, \mathrm{Mpc}} \right) + 25
\label{eq:mu_sne}
\end{equation}
where $z$ is the redshift, $\theta$ are the cosmological parameters  and $D_{\rm L}$ is given by 

\begin{equation}
    D_{\rm L} = \frac{c (1 +z)}{H_0}\int_0^z\frac{dz}{E(z)}\, .
\end{equation}

We can write the dimensionless Hubble parameter, $E(z)$ for the evolution as
\begin{equation}
    E^2(z) =  [\Sigma_x \Omega_x a^{-3 \cdot (1 + w_x)}]
\label{eq:hz}
\end{equation}
where $a$ is the scale factor related to the redshift as $a = 1 / (1+z)$. For a generic time-varying equation of state (see e.g., \citet{2001A&A...380....6G}) one can substitute
\begin{equation}
    a^{-3 (1 + w_x)} \rightarrow {\rm exp} \left( 3 \int_a^1 \frac{1 + w(a)}{a} da \right)
\end{equation}
  We will focus on the dark energy model given by the CPL parametrisation, i.e.
 \begin{equation}
     w(a) =  w_0 + (1-a) w_a
 \end{equation}
 which has most commonly been used to test the standard model (however see section~\ref{sec:nonstan_models} for an exploration of other parametrisations). 

Observationally, the distance modulus is calculated from the SN~Ia peak apparent magnitude ($m_B$), light curve width ($x_1$) and colour ($c$):
\begin{equation}
\mu_{\rm obs} = m_B - (M_B - \alpha x_1 + \beta c) + \Delta_\mathrm{M} + \Delta_\mathrm{B},
\label{eq:obs_distmod}
\end{equation}

where $M_B$ is the absolute magnitude of an SN~Ia with $x_1 = c = 0$. Here, $\alpha$, $\beta$, are the slopes of the width-luminosity and colour-luminosity relation, and are fit to the whole sample, in contrast to $m_B$, $c$, and $x_1$, which are fit per-SN. $\Delta_\mathrm{M}$ and $\Delta_\mathrm{B}$ are  host galaxy luminosity corrections and distance bias corrections respectively \citep[see, ][for details ]{Popovic2021, Popovic2023}. We generate simulated $\mu_{\rm obs}$ from an input cosmology with ($\Omega_{\rm M}$, $w_0$, $w_a$) = (0.3, -1, 0) - i.e. the same input cosmology as for the BAO and CMB datasets. To simulate the intrinsic dispersion in the SN~Ia magnitudes we scatter the distance moduli with a $\sigma_{\rm int}$ of 0.12 mag, which is the typical intrinsic scatter of SNe~Ia, \citep[e.g.][]{Foley2018,Brout2022_cosmo}.

\begin{figure}
    \centering
    \includegraphics[width=0.5\textwidth]{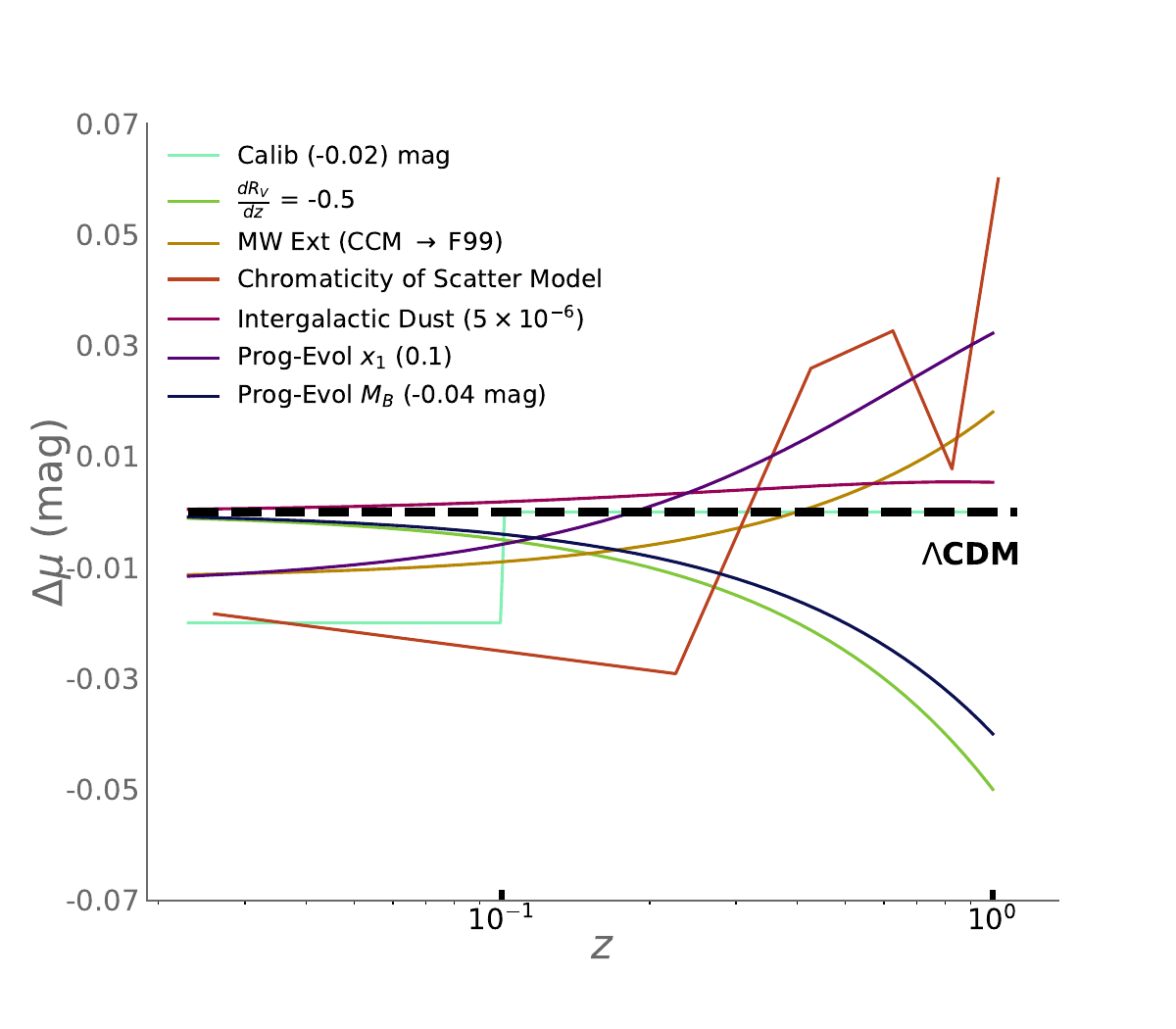}
    \caption{Changes to the Hubble Residuals from the input cosmological distances caused by the systematic uncertainties analysed in this work. The description of each effect is summarised in Section~\ref{sec:sys-group}; we not that the amplitude and direction for each systematic uncertainty is chosen for ease of visualisation, with the exception of the Intrinsic Scatter Model and MW Dust Law. For the chromaticity of the intrinsic scatter model, we have smoothed the curve with a spline function for visual clarity. All of the systematic effects can act over a wide range of values that are accessible in the python program provided in this work; the ranges are discussed in  Section~\ref{sec:sys-group}. We do not show the $\Omega_{\rm M}$ mismatch here, as its impact is not visible at the Hubble Residuals level. }
    \label{fig:residuals}
\end{figure}

To infer the cosmological parameters, we need the full SN~Ia covariance matrix $C_{\rm SN}$, which is given by
\begin{equation}
    C_{\rm SN} = C_{\rm Stat}  + C_{\rm Sys}
\label{eq:c_sn}
\end{equation}
where $C_{\rm stat}$ is the statistical covariance matrix and $C_{\rm sys}$ is the systematic covariance matrix.
Since the likelihood for the dataset is Gaussian, we only need to sample over the input parameters and fit the $\chi^2$ distribution. The $\chi^2$ is given by 
\begin{equation}
\chi_{\mathrm{SN}}^2 = \Delta^T C_{\mathrm{SN}}^{-1} \Delta, 
\end{equation}
where $\Delta = \mu- \mu_{\rm obs}$ and $C_{\mathrm{SN}}$ is the complete covariance matrix described in \cite{Brout2022}.  

We sample over the parameters and compute the likelihood using a nested sampling algorithm, \texttt{MultiNest} as introduced in \cite{Feroz2009,Feroz2019}, via the python module \texttt{pymultinest} \citep{Buchner2014}. We note that for both the SN~Ia magnitude-redshift relation and the BAO+CMB ``external data" combination, we assume that since the underlying cosmology is the same, the input cosmological model for the simulated observables is the same. The impact of any mismatch between the inferred cosmology from the two probes is studied in Section~\ref{sec:impact_de}.

\begin{figure}
    \centering
    \includegraphics[width=0.5\textwidth]{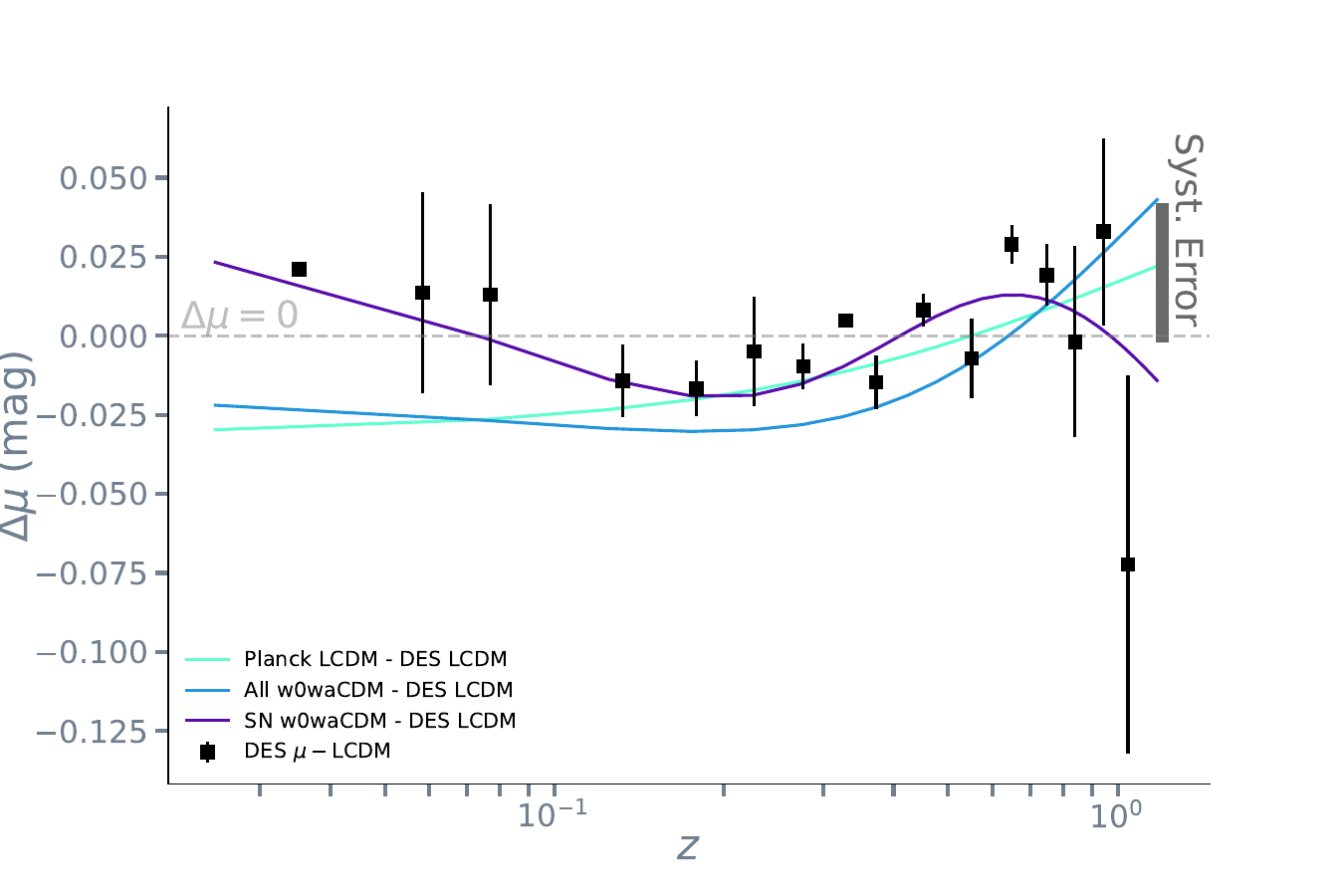}
    \caption{Residuals of the DES-5YR SN~Ia sample \citep{Vincenzi2024,Sanchez2024} relative to the best-fit $\Lambda$CDM cosmology  (violet), compared with the residuals of the best fit CPL cosmology to $\Lambda$CDM (black). The Planck (+BAO) data is used in combination with the DES-5YR SNe~Ia for the best fit cosmology to the SBC combination, which is shown as the cyan line. The typical amplitude of the systematic uncertainties tested here is shown as the solid blue line, which is comparable to the difference in the residuals for the very different cosmologies shown in the dotted lines. This illustrates how sensitive the cosmological inference is to the systematic uncertainties.  }
    \label{fig:hubres_cosmo}
\end{figure}
\begin{figure*}
    \centering
    \includegraphics[width=0.48\textwidth]{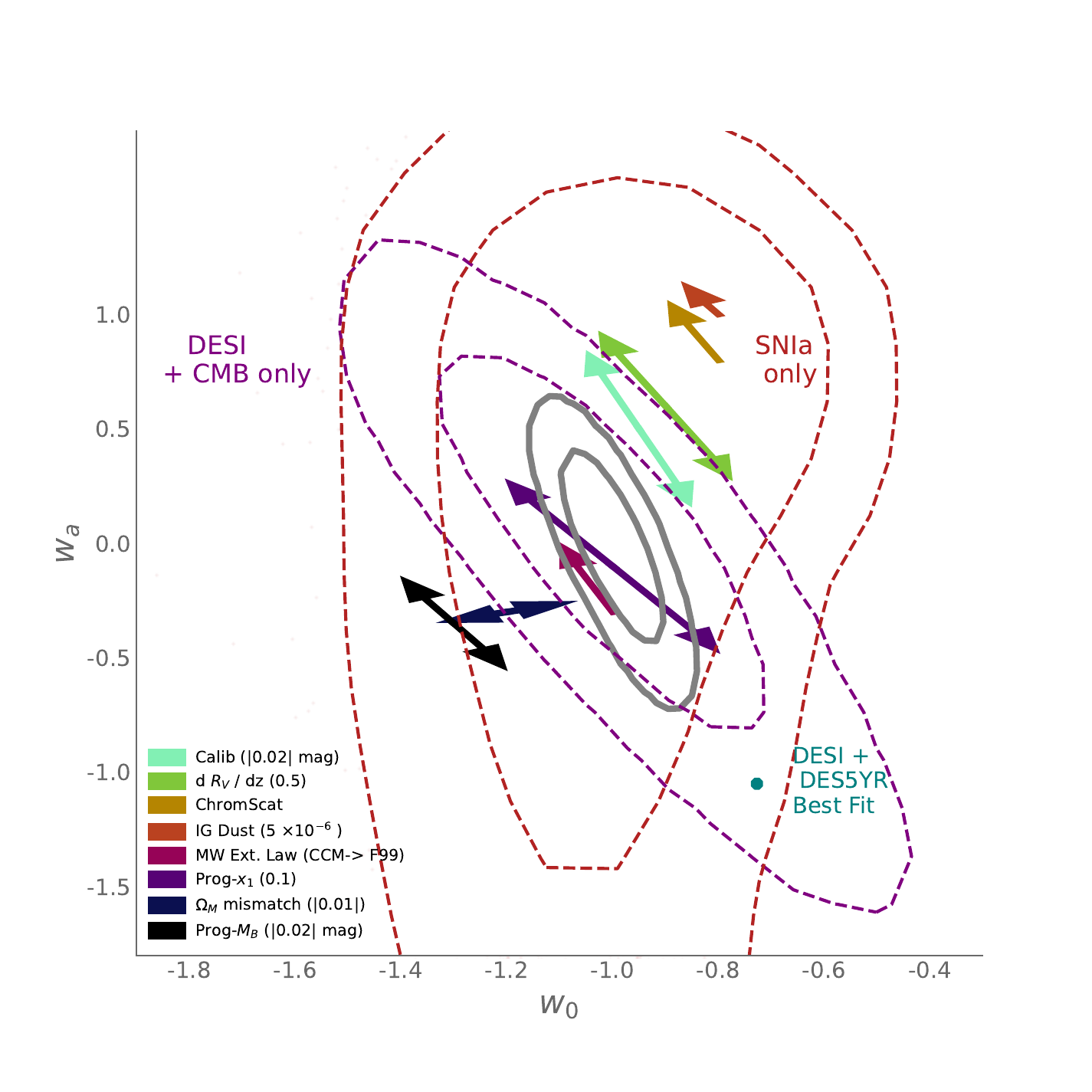}
\includegraphics[width=.48\textwidth]{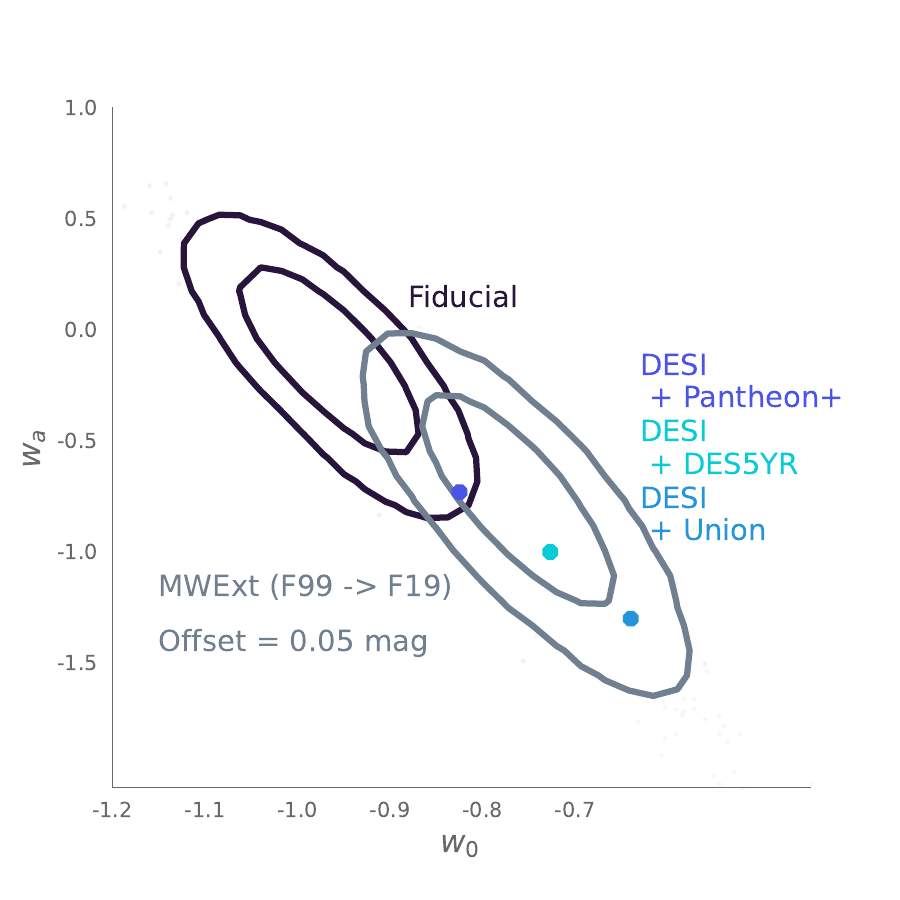}
    \caption{(Left): 68$\%$ and 95$\%$ confidence intervals for the BAO+CMB (purple), SN~Ia (red) and combined (gray) probe combinations.A summary of the systematic grouping effects and the respective shift in the $w_0 - w_a$ plane. The amplitude and direction of the arrows correspond to the deviation of the $w_0$-$w_a$ values from the fiducial cosmology of (-1. 0). For effects where the bias on the distance can be scaled, we show bi-directional arrows since those effects can take both negative and positive values, e.g. calibration offset can be either positive or negative for low-$z$ compared to high-$z$ and similarly with the change in dust with redshift. For comparison the best fit value from DESI+DES5Yr+CMB is shown in teal. (Right): The shift in the inferred contours from the fiducial (black) cosmology to the SBC combined result from SBC combination with Pantheon+, Union and DES-5Yr SN~Ia samples (while the best fit from the Union compilation looks the most deviant from (-1, 0) it has larger errors than DES5-Yr). This can be explained by a difference in calibration of 0.05 mag and a shift in the extinction law used for Milky Way reddening correction.}
    \label{fig:summary_systematic} 
\end{figure*}
\section{Systematics Groupings}
\label{sec:sys-group}
Here we summarise the different themes for systematic errors that are tested in this study. We explore experimental and astrophysical sources of error that can alter the distance modulus as a function of redshift. The input functional forms and individual values are summarised in each subsection. We note that each of the effects is analysed assuming that the data is already corrected for selection effects. A comparison of the residuals of the best fit cosmological model to the average amplitude of the systematic error is shown in Figure~\ref{fig:hubres_cosmo}

\input{tables/sys_delt_tab}

\subsection{Calibration}
Historic SN Ia survey designs have precluded a simultaneous collection of high-$z$ and low-$z$ data on a single telescope; therefore, cross-calibration of telescopes to ensure consistency has been crucial step in measuring the properties of dark energy. Historically this cross-calibration has been performed with field stars (\citealp{Brout2022,Vincenzi2024}, Popovic et al. 2024d \textit{in prep.}). Calibration issues have been exacerbated by the existing low-redshift sample being a collection of multiple telescopes and systems that, in some cases, no longer exist. For this reason, there are several low-$z$ SN Ia programs to observe a uniform sample of SNe Ia on a single, well-characterised photometric system
\citep{Foley2018,Dhawan2022,Rigault2024}. 
However, even these improved and uniform low-$z$ systems will require cross-calibration with high-$z$ data sets. Here, we model the impact of calibration as an offset between the SNe~Ia at $z \leq 0.1$ and $z > 0.1$. We allow for an offset between our low- and high-$z$ data that can range from an optimistic 0.01 magnitudes to a highly conservative 0.1 magnitudes. Since the low-$z$ can be miscalibrated to be either brighter or fainter relative to the high-$z$ we analyse $w_0 - w_a$ a range of inputs from -0.1 to 0.1 mags.

\subsection{Milky Way Extinction}
The use of SN Ia lightcurves for precision cosmology requires correcting for dimming and reddening caused by dust within the Milky Way galaxy. This is performed by assuming a dust law with a specific extinction ratio $R_V = 3.1$ and a colour extinction $E(B-V)$ from Galactic dust maps from \citet{Schlafly2011} (hereafter, SF11). Typically, the dust law used is \citet{Cardelli1989}, hereafter CCM89. However, there have been updates to the galactic dust law e.g., \citet{Fitzpatrick1999} (F99), \citet{Fitzpatrick2019} (F19).  In F19, 72 UV-optical-NIR (UVOIR) extinction curves are computed from IUE and HST/STIS UV-optical spectro-photometry (from 1150 \AA\,to 10000 \AA), combined with NIR photometry. Even for the same $R_V$ there are differences between the F19 treatment of the extinction curve compared to previously computed dust laws \citep[see also;][]{Mortsell2022}. The
high-resolution curves also allow the study of extinction features on intermediate wavelength scales in the optical with a high S/N. We, therefore, explore how for a fixed $R_V$ the dust law itself can change the inferred apparent peak magnitude, and hence, the distance.  
For this systematic effect, we test the difference in the predicted $m_B$ between the difference dust laws. We extinguish the SALT3 SN~Ia SED in the observer frame with the MW dust law for given $R_V=3.1$ and an $E(B-V)_{\rm MW}$ from the SF11 map corresponding to the coordinates of observed SNe~Ia from the Pantheon+ compilation. We emphasize that the specific sky distribution is not important as long as it probes a significant number of different sightlines, like the Pantheon+ compilation does. We take the CCM89 dust law as the reference point. For our simulated Hubble diagram, we pick the $E(B-V)_{\rm MW}$ as a function of the coordinates of the SN in the simulation and the SF11 map. For both the F99 and F19 dust laws, we simulate the systematic difference in the distance modulus as the difference in the predicted $m_B$ from that dust law and CCM89. The example of the difference is plotted in the top panel of Figure~\ref{fig:residuals} termed as the ``MW-Ext" curve. The impact on $w_0$ and $w_a$ is summarised in Table~\ref{tab:sys_delt}.

\subsection{Chromaticity of Intrinsic Scatter}
Modelling the intrinsic scatter of SNe~Ia as a function of redshift is a non-trivial task. Improvements in our knowledge of the dust distribution in the low- and high- mass host galaxies and the dependence of the Hubble residuals on the SALT2 colour was seen to improve the distance bias correction \citep{BS21} (hereafter BS20) one of the key intrinsic scatter models adopted in the Pantheon+ analysis. More recent models, e.g. \citet{Popovic2023}  (hereafter P23) - which simultaneously compute the scatter as a function of the SN  colour and dust parameters,  have been used in the fiducial DES-5yr analysis. While the difference with other intrinsic scatter models has been factored as a source of systematic in the error covariance matrix, here we explore what impact the explicit difference between the two models has on the $w_0-w_a$ inference. To first order, such a difference can be thought of as an improvement in the inference of the dust population parameters and hence, in the intrinsic scatter arising from a ``chromatic" term. Hence, this systematic effect is called ``ChromScat". 
Here, we use the difference in the predicted bias between the BS20 and P23 models as the systematic corresponding to the intrinsic scatter, shown as the black curve in Figure~\ref{fig:residuals}. The impact on $w_0-w_a$ is shown in Table~\ref{tab:sys_delt} and Figure~\ref{fig:summary_systematic}. 

\subsection{Intergalactic Dust}
While the SED model fit to the broadband data of SNe~Ia takes into account the dust from the host galaxy either as an empirical colour term or an absorption in the $V$-band, the current cosmological analyses do not include the absorption from intergalactic dust, $\Omega_{\rm dust}$. While the current best constraints on the total intergalactic dust density budget show that it is a small fraction of the total baryonic matter \citep[e.g.,][]{Goobar2018}, we test what impact that can have on the inferred $w_0$-$w_a$ constraints. 
We summarise the computation of the bias on the Hubble residuals ($\Delta$) from the intergalactic dust below \citep[see][for details]{Goobar2018}
The optical depth for the scattering from intergalactic dust is given by

\begin{equation}
    \tau = \kappa_\lambda\rho_{\textup{dust}} d
\end{equation}
$d$ is the light-travel distance and $\rho_\textup{dust}$ the density of the dust and $\kappa$ is the optical depth. The dust distribution is assumed to be homogeneous throughout space, but not in time (redshift). This assumption is justified as dust presence has been detected at distances of several Mpc outside galaxies \cite{Menard2010}. $\tau$ can be expressed as 
\begin{equation}
    \tau = \int \kappa_\lambda\rho_{\textup{dust}}(z)\cdot cdt = \int_0^{z_s} \kappa_\lambda\rho_{\textup{dust}}(z)\cdot c\frac{dt}{dz}dz
\end{equation}
where $z_s$ indicates the redshift of the source. Note that
\begin{equation}
    \frac{dt}{dz} = \frac{1}{(1+z)H(z)}
\end{equation}
Since $\kappa$ is a function of $\lambda$, it would be redshift dependent. We assume, $\kappa_v \sim 1.54\times 10^{4} {\rm cm}^2 {\rm g}^{-1}$ \citep{Weingartner2001}. If we observe $\lambda_0$, the light was emitted with a wavelength $\lambda_{\textup{emit}}=\frac{\lambda_0}{1+z_s}$. The wavelength as a function of $z$ is thus
\begin{equation}
    \lambda(z) = \lambda_{B}\frac{1+z_s}{1+z}
\end{equation}
where refers to the effective wavelength of the $B$-band in the rest frame. We treat the IG dust, similarly to other cosmological components, i.e. parametrised by some value of its current density and a function describing its redshift dependence, parameterized by some exponent $\gamma$.
\begin{equation}
    \rho^{\textup{dust}}(z) = \rho^{\textup{dust}}_0\cdot (1+z)^\gamma
\end{equation}
Scale it with the critical density 
\begin{equation}
    \rho^{\textup{dust}}(z) = \frac{3H_0^2}{8\pi G}\Omega_{\textup{dust}}(1+z)^\gamma
\end{equation}

\begin{equation}
    \tau_\lambda = \frac{3cH_0}{8\pi G}\Omega_{\textup{dust}}\int_{0}^{z_s}\frac{\kappa\left ( \lambda \left (\frac{1+z_s}{1+z}\right ), R_V\right )\cdot (1+z)^{\gamma-1}}{E(z)}dz
\end{equation}
Expressed as a difference in magnitudes
\begin{equation}
   \Delta_{\rm HR} = -2.5\log_{10}e^{-\tau_\lambda} \approx 1.086\tau_\lambda
\end{equation}

For our tests, we take the dust density in steps of log($\Omega_{\rm dust}$) from $10^{-10}$ to $10^{-4}$. 

\subsection{Progenitor Evolution: $x_1$}
The observed properties of SNe~Ia can depend on progenitor properties, e.g., metallicity \citep{MR2016a,MR2016b}, age \citep{Childress2014}. Since these progenitor properties can evolve with redshift, it can have potential impact on the intrinsic luminosity and lightcurve properties of the SN.  Such an evolution can manifest in the evolution of the $x_1$ distribution from low- to high-$z$. An effect of such kind is particularly important, given the recent findings that the width-luminosity relation correction itself maybe different for the low- and high-$x_1$ end of the distribution \citep{Ginolin2024}.   We take the simple case where the $z$-dependence of $\alpha$ is an outcome of the dependence of $\alpha$ on $x_1$ via a broken power law. We use an $\alpha_{\rm low} = 0.23$ and $\alpha_{\rm high} = 0.13$ and where low is $x_1 < -0.49$ and high is $ >= -0.49$. We assume, for the simplistic case that the selection effects are corrected for and that the mismatch in the true and inferred distances comes from correcting a redshift evolving $x_1$ distribution which has an $x_1$ dependent $\alpha$ using a single $\alpha$ that is $x_1$ independent. Details of the computation are described in section~\ref{sec:prog_x1_app}. The resulting Hubble residuals are plotted in Figure~\ref{fig:residuals}. 
\subsection{Progenitor Evolution: $M_B$}
\label{sec:prog_mb}
Spectroscopic observations of intermediate ($z \sim 0.5$) and high ($z > 1$) SNe~Ia have shown a remarkable similarity in the optical \citep{Balland2009,Balland2018,Dhawan2024b}, and to varying degree in the ultraviolet \citep{Ellis2008,Maguire2012,Foley2012}. While the spectroscopic similarity would indicate no strong evolution of SN~Ia properties with redshift, subtle evolution in the brightness of the population has not been ruled out. 
We, therefore, parametrise the evolution of the progenitor properties in form of an evolving intrinsic luminosity as a simple, linear, function of the redshift. The slope of this evolution can intuitively be interpreted as a difference between the average luminosity of the sample at $z = 0$ and $z = 1$.  We take a range of slope values from -0.1 to 0.1, such that we would include both cases where the SNe~Ia at $z = 0$ are brighter than $z = 1$.  

\subsection{$\Omega_{\rm M}$ mismatch}
Currently, the constraints on $w_0$-$w_a$ are reported from a combination of early and late universe probes, i.e. the CMB, BAO and SNe~Ia. We note that in a flat $\Lambda$CDM model, i.e. a cosmology with $(\Omega_K, w_0, w_a) = (0, -1, 0)$, the predicted $\Omega_M$ from the Planck satellite ($0.315 \pm 0.007$) is lower than the measurement from the DES SNe~Ia ($0.357 \pm 0.017$), at the $\sim 2.5 \sigma$ level. Owing to the degeneracy between $\Omega_M$, $w_0$ and $w_a$ when the CMB and BAO data is combined with SNe~Ia, it can shift the inferred contours, compared to the case where the two probes would indicate the same value of $\Omega_M$. We simulate this effect as a mismatch between the $\Omega_M$ input for simulating the SN~Ia Hubble diagram and the value used as input for the external data. We note that while such a mismatch can be due to unresolved systematics in any of the categories described above, we represent it in the form of the $\Omega_M$ parametrisation as an ``unknown unknown", i.e. an effect that is likely to impact the $\mu - z$ relation but we have not deviced a parametric form for it.

\subsection{Increased scatter in colour correction}
While systematic offsets can affect the relative distance modulus between low-$z$ and high-$z$ SNe~Ia, a dispersion in the distance inference - e.g. due to diversity in intrinsic or extrinsic properties - can also increase the uncertainties without changing the central value. Here, we analyse how the inferred constraints on $w_0 - w_a$ change with increased scatter in the colour corrections. The Tripp formula is used to correct the apparent peak luminosity. For this correction, a single colour-luminosity correction is assumed, i.e. a $\beta \times c$ term. This can be seen as two separate corrections, one for the intrinsic colour $\beta_{\rm int} c_{\rm int}$ and the other for the dust law $R_V E(B-V)$. However, we can expect that there is some dispersion in this correction, either due to unaccounted for differences in the slope of the dust law or differences in the intrinsic colour-luminosity relation. We dub this term as a `$\sigma(\beta)$', systematic. 
For our analysis, we treat this term as a simple increase in the diagonal terms of the covariance matrix.  We take a range of $\sigma(\beta)$ values corresponding to a constant error term of 0.15 mag. While we interpret this additional term as a colour correction term, it can be generalised to any other scatter term in the distance modulus, e.g. a residual intrinsic scatter. 

\section{Impact on dark energy}
\label{sec:impact_de}
In this section, we summarise the expected constraints from the combination of low- and high-$z$ SNe~Ia, combined with priors from external data,  in the $w_0$-$w_a$ plane. We test each of the systematics groupings with the varying amplitude of the effect in each subsection below. A user interface to test the different effects with varying amplitude is provided as public software along with this paper \footnote{\hyperlink{here}{https://github.com/sdhawan21/DEslider   }}.  We find that the shift in $w_0 - w_a$ for all the systematics groups is close to the orientation  of the $w_0 - w_a$ degeneracy (Figure~\ref{fig:summary_systematic}).  

\subsection{Dark energy constraints with systematics}

We test the impact of residual systematics - i.e. effects that are not corrected for after the treatment of the lightcurve fit parameters and hence, distances - on $w_0$-$w_a$.  We find that most of the systematic effects tested here align with the axis of the $w_0$-$w_a$ constraints and in the direction of the shift from the input cosmology towards the DESI results. For the first effect of photometric calibration, we find that an offset of 0.02 mag can shift $w_0 - w_a$ by $\Delta w_0$  of 0.073 and $\Delta w_a$ of -0.248 as demonstrated by the light blue arrow in figure~\ref{fig:summary_systematic}. For an evolving dust population such that the mean $R_V$ is higher by 0.5 at $z = 1$ compared to the low-$z$ (i.e. $z \sim 0.05$) universe the constraints can shift by $\Delta w_0= -0.092$ and $\Delta w_a = 0.234$
We also find that using a more updated dust law for the MW correction can shift the contours by $\Delta w_0= -0.071$ and $\Delta w_a = 0.211$. We compute this relative to the commonly used galactic extinction law F99. However, when comparing to a more updated MW dust law, F19, the shift is $\Delta w_0= -0.077$ and $\Delta w_a = 0.157$. 

For the intergalactic dust component, we present the difference in units of log($\Omega_{\rm dust}$) and find that there is only a significant shift in $w_0$-$w_a$ if $\Omega_{\rm dust} \gtrsim 5 \times 10^{-6}$. For  $\Omega_{\rm dust} = 5 \times 10^{-6}$, we find a $\Delta w_0 = -0.027$ and $\Delta w_a = 0.052$.

The model of the intrinsic scatter - i.e. swapping the fiducial scatter model for one with a different fraction of scatter being chromatic - can shift the constraints by $\Delta w_0 = -0.063$ and $\Delta w_a = 0.168$
Testing for progenitor evolution as presented in \citet{Nicolas2021}, we find a shift in $\Delta w_0= -0.157$ and $\Delta w_a = 0.292$. When using a linear evolution of $M_B$, we find dark energy parameter shift by $\Delta w_0= -0.058$ and $\Delta w_a = 0.116$. 
A summary of the size and direction of the shift in $w_0-w_a$ is shown in Figure~\ref{fig:summary_systematic}. Three effects, namely, ChromStat, MWExt and IG Dust only have single direction arrows. For the first two, this is because we test the difference between the fiducial assumption and an alternate model and hence, the direction is set by the choice in the default SBC analysis \citep{DES2024,DESI-VI}. If the choice were to be inverted (e.g. if the dust law assumed was F99 and we tested the impact of changing it to CCM89), it would also invert the arrow, but we show only the direction relative to the fiducial case from the DES 5Yr analysis in the Figure. For the IG Dust systematic, since dust only dims the SNe, the effect would only move the $w_0-w_a$ contour in one direction.  

\vspace{-0.05cm}
\subsection{Comparison to current results}
From the previous section, we can see that there are several effects that can significant impact on dark energy inference in the $w_0$-$w_a$ plane. The orientation of the systematics shifts presented in Figure~\ref{fig:summary_systematic} and is closely aligned with the direction of the deviation reported in \citet{DESI-VI} relative to $\Lambda$CDM, i.e. (-1, 0). We illustrate what combination of systematic groupings and their associated amplitudes can cause a shift in the inferred $w_0$ - $w_a$ away from the (-1, 0) - i.e. $\Lambda$CDM - case to the combination of SBC in \citet{DESI-VI}, i.e. $w_0 = -0.725$ and $w_a = -1.06$. 
An example case illustrated in Figure~\ref{fig:summary_systematic} shows a combination of systematics from calibration offset and a change in the galactic extinction. We find that a calibration offset of 0.05 mag i.e. such that the SNe at $z < 0.1$ are fainter by 0.05 mag than the $z \geq 0.1$ sample and an difference between MW extinction using F99 and F19 dust laws can shift the central value of the $w_0$-$w_a$ contours such that the best fit DESI value is well within 1-$\sigma$ when the true input cosmology is $\Lambda$CDM.

\begin{figure}
    \centering
    \includegraphics[width=0.5\textwidth]{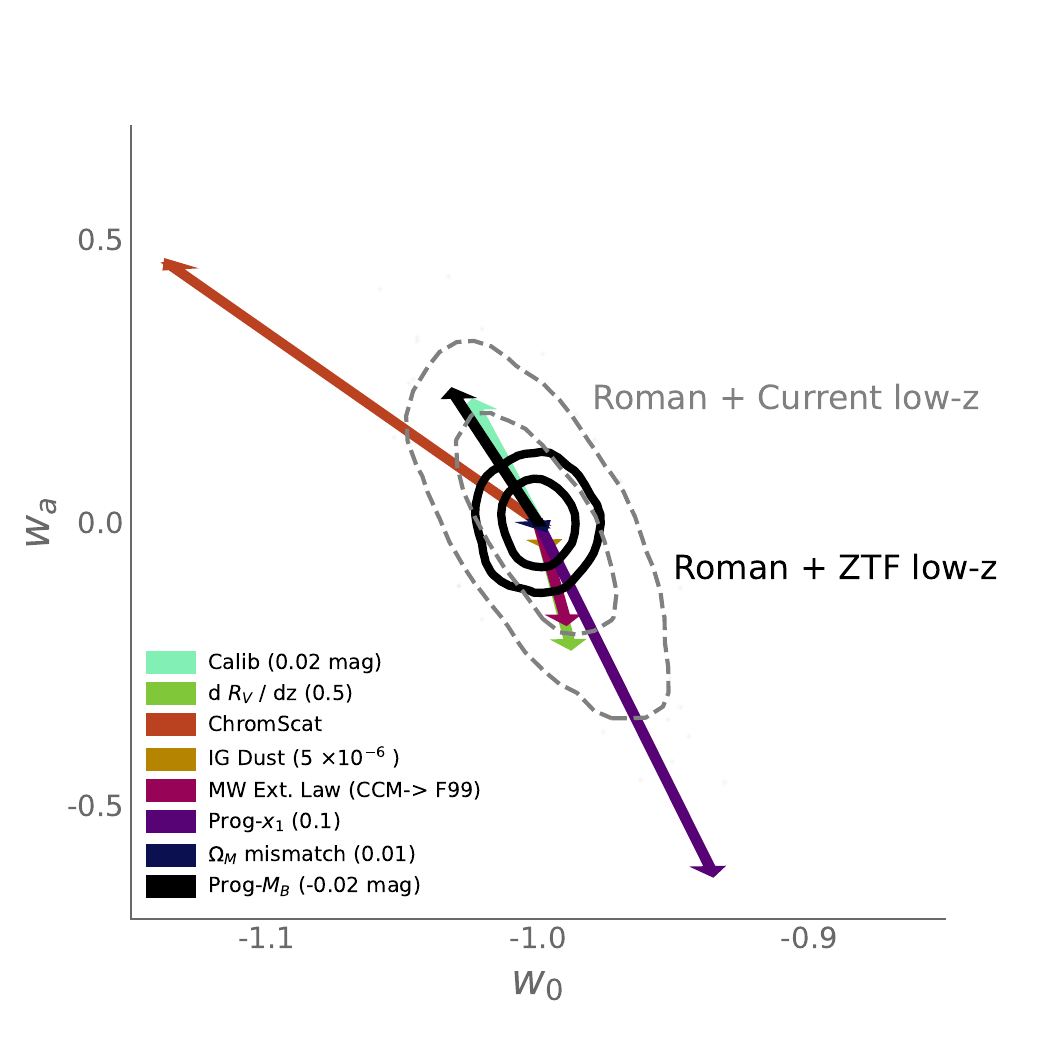}
    \caption{68$\%$ and 95$\%$ confidence intervals and shifts in the $w_0$-$w_a$ plane for the expected SN~Ia dataset from the Roman Space Telescope + ZTF (black) and the complete set of DESI BAO measurements. The shifts follow different degeneracy directions in the $w_0 - w_a$ plane. For comparison, the $w_0-w_a$ constraints expected if the low-$z$ anchor comprises the present data dataset is shown in grey. From this figure, it is evident that with next generation low- and high-$z$ SN~Ia surveys, the systematic uncertainty will need to be significantly reduced to robustly estimate $w_0-w_a$. }
    \label{fig:summary_future}
\end{figure}

\section{Discussion and Conclusions}
\label{sec:discussion_conclusions}
We analysed  how different types of systematic errors in SN~Ia analyses \citep[e.g.][]{Scolnic2014,Vincenzi2024} can impact the inferred cosmological parameters describing the properties of dark energy. We found that a mismatch in calibration, change in the dust law with redshift, chromaticity of the intrinsic scatter, evolution of the progenitor properties and the assumption of the galactic dust law can shift the inferred cosmology from the true input parameters. Crucially, the shift in the $w_0$-$w_a$ from these effects is along the direction of the best fit cosmology from combined SBC probes \citep{DESI-VI,DES2024}. For a small shift in the calibration of 0.05 mag and a change in the MW dust extinction law, we see a shift from a true $\Lambda$CDM cosmology to the best fit value from the SBC probe combination. 
We emphasize that in this work, we take the BAO+CMB data covariance without any additional systematics perturbing those distances. Since the BAO+CMB are very constraining in the $w_0$-$w_a$ plane, any residual systematics in the BAO+CMB probe combination could also impact the degeneracy direction of the shifts. 


\subsection{Additional Systematic Uncertainties}
While the groups of systematic uncertainties explored here cover a wide range of astrophysical and instrumental effects, the list is not exhaustive. Other sources of uncertainty can include effects like weak lensing from the large scale structure of the universe \citep[e.g., see][]{2007JCAP...06..002J,2014ApJ...780...24S,Shah2024}. Conservatively, this systematic uncertainty term has been considered in the literature as a linear function of the SN~Ia redshift, most commonly as $\sigma_{\rm lens} = 0.055 z$ \citep{Holz2005}. Such a systematic has an identical effect on the dark energy inference as a linear evolution of $R_V$ as a function of redshift\footnote{This effect can also be replicated in the accompanying widget using the ``prog-evol-mb" dial.}. Hence, the impact on dark energy for different amplitudes of the weak lensing signal can be evaluated similar to the dust evolution effect, using the tools provided with this paper.

Another possible residual systematic is the misclassification from photometry of core collapse SNe  as SNe~Ia\citep[e.g., as discussed in][]{Vincenzi2024}. This could impact the intrinsic  luminosity in different redshift bins differently, because the CCSNe being intrinsically fainter could impact the lower-$z$ bins, however, the higher-$z$ bins will likely  have less contamination as CCSNe would not be detected. This effect can be approximated as a linear evolution of $M_B$ as in the progenitor evolution systematic described in section~\ref{sec:prog_mb}.
A summary of the impact of the different systematics is shown in Figure~\ref{fig:summary_systematic}. 

\subsection{Impact of systematics for future surveys}
We presented the expected shift in the $w_0$-$w_a$ constraints from the combined SBC probes. The inference is based on the error budget from current surveys, to match the error ellipse in \citet{DESI-VI,DES2024}. However, we expect that both SN~Ia and BAO data will improve significantly in the coming years with current and oncoming stage-IV surveys. We, therefore, analyse how these shift will differ for future datasets. We perform the same analysis with the forecast for the Roman Space Telescope SNe~Ia magnitude-redshift relation \citep{Hounsell2018} at high-$z$, complemented by a low-$z$ anchor from ZTF  - corresponding to the size of the current DR2 sample \citep{Rigault2024}. For the external data we take the complete DESI forecast \citep{desi-forecast-2016} and since {\it Planck} has measured the first peak of the CMB power spectrum with extreme precision, we keep to using the {\it Planck} 2018 compressed data errors for the CMB constraints. While other observatories will also measure SN~Ia and BAO distances \citep{Laureijs2011,lsst_srd}, we using the Roman+DESI combination since the forecast involves realistic simulations. The resulting shift for each of the systematic uncertainties is shown in Figure~\ref{fig:summary_future}. As the next generation datasets reduce the error in the $w_0 - w_a$ plane, the systematic uncertainties can shift the contours by $\sim 2-3\sigma$ and for certain effects $> 5 \sigma$ (compared to only 1 - 2$\sigma$ shifts we see in Figure~\ref{fig:summary_systematic}). From Figure~\ref{fig:summary_future}, it is also evident that an improved low-$z$ anchor sample of 2500 SNe~Ia \citep[e.g.,][]{Rigault2024} can improve constraints on $w_0-w_a$ by a factor of $\sim 3$.  From our simulated effects, we find that progenitor evolution and the chromaticity of intrinsic scatter are the two most important systematics, however, the change in dust population and calibration can also lead to shifts that are of order the size of the error ellipse. While the specifications of future surveys are likely to change, given the detail in the simulations, the scale of these shifts can be considered realistically representative.

In our work, we focussed on the canonical CPL parametrisation to compare to the inference in \citet{DESI-VI,DES2024} and to ascertain the key systematic effects to be controlled for future missions, which are designed based on their ability to improve constraints in the CPL model. However, there are other approaches to model accelerated expansion (see also, Appendix~\ref{sec:nonstan_models}) within which future work can explore the role of systematics \citep{Lovick2023,Mukhopadhayay2024}.
Therefore, for current and future Stage-IV missions, it is extremely important to control the systematic uncertainties for robustly testing for deviations from $\Lambda$CDM.  

\section*{Acknowledgements}
We thank George Efstathiou for interesting discussions. SD acknowledges funding from a Kavli Fellowship and a JRF at Lucy Cavendish College. AG acknowledges support from the Swedish Research Council and the Swedish National Space Agency. 

\section*{Data Availability}
The interactive widget along with this study is available publicly via github here:\hyperlink{here}{https://github.com/sdhawan21/DEslider.git} . 
 



\bibliographystyle{mnras}
\bibliography{example} 

\appendix
\section{Alternate dark energy parametrisations}
\label{sec:nonstan_models}

\begin{figure}
    \centering    \includegraphics[width=0.48\textwidth]{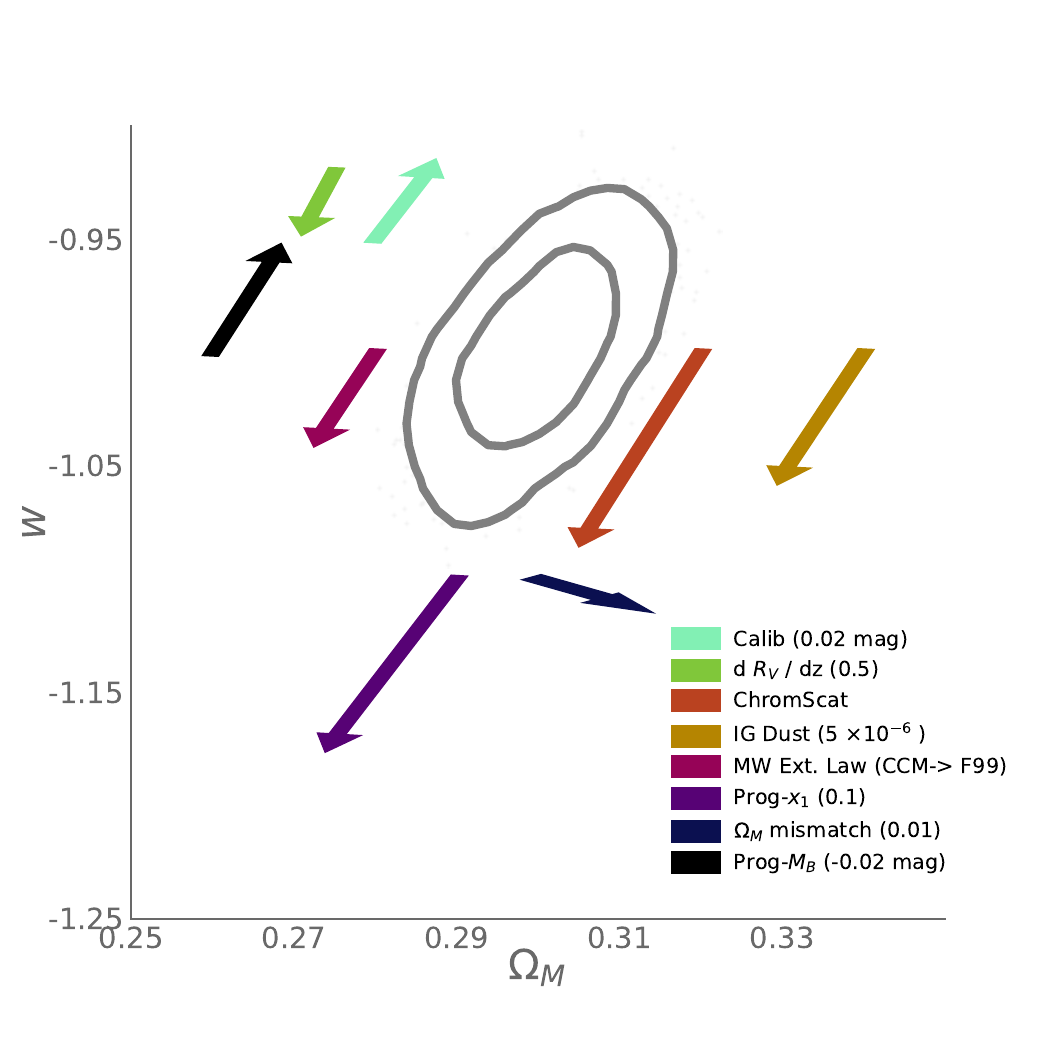}
    \caption{Constraints and the direction in which each of the systematic effects deviates the result from the fiducial case for $w$CDM. We find that the systematic effects (apart from the $\Omega_{\rm M}$ mismatch all impact the cosmology constraints along the degeneracy direction of smaller $\Omega_{\rm M}$ and more negative $w$.}
    \label{fig:summary_wcdm}
\end{figure}

In this paper, we have explored a large range of systematic uncertainty groups and their impact on the present day equation of state of dark energy ($w_0$) and its time-dependence ($w_a$) in a CPL model. While this model is most commonly used to test the deviations from $\Lambda$CDM, there are other parametrisation of the dark energy equation of state and physical descriptions of accelerated expansion that can be tested \citep[e.g.][]{Lovick2023,Calderon2024,Mukhopadhayay2024}. We test a simple phenomenological model with a constant $w$ and also explore the impact on physically motivated dynamical scalar field models.  
In this case, the $H(z)$ is given by
\begin{equation}
    H^2/H_0^2 = \left[\Omega_{\rm M} (1+z)^3 + (1 - \Omega_{\rm M})(1+z)^{3 (1+w)} \right]
\label{eq:hz_wcdm}
\end{equation}
We summarise the impact of the different systematic groupings in the constant $w$ space in figure~\ref{fig:summary_wcdm}. . The effects of progenitor evolution and change in $R_V$ bias $w$ to more negative, i.e. if the SNe~Ia are 0.04 mag fainter, then the inferred $w$ is more negative by 0.022 (i.e. -1.022 instead of the true value of -1). In case of the calibration, the high-$z$ SNe~Ia are 0.02 mag brighter than the low-$z$ making the inferred $w$ biased less negative by 0.028.

While the CPL model is powerful to test deviations from $\Lambda$CDM, it is purely phenomenological. One physical explanation for accelerated expansion is that it is driven by a dynamical scalar field, one which is frozen to an equation of state close to -1 in the early universe and is thawing from it \citep[the thawing / freezing nomenclature was introduced in][]{Caldwell2005}. We can expect there to be a diverse range of functional forms of the potential of the scalar field $V(\phi)$. Some examples include a pseudo-Nambu-Goldstone Boson \citep{Frieman1995,SmerBarreto2017}, an algebraic thawing model \citep{Linder2015}. For our analyses we test a formalism that generalises the equation of state as a function of redshift for these classes of models.  The equation of state for these models is given as 
 \begin{equation}
     w(z) = -1 + (1 + w_0) e^{-\alpha z}
 \end{equation}
 where $w_0$ is the present day equation of state and $\alpha$ characterises a diverse set of families of thawing quintessence models while having a narrow distribution of 1.45 $\pm 0.1$ \citep[see][for details]{Camilleri2024}. We find that the systematic groupings shift $w_0$ very similarly to $w$ in the $w$CDM model. 

\section{Progenitor evolution via $x_1$ population drift}
\label{sec:prog_x1_app}
We note that while several possible physical causes of diversity can lead to an evolution in the $x_1$ distribution, a prescription for 
 the $x_1(z)$ distribution is given in equation 2 of \cite{Nicolas2021} and the $\delta(z)$ which is a function of LsSFR is in Eq. 1 of that paper. 
 The differential with respect to the underlying true cosmology is given by 
\begin{equation}
    \Delta_{\rm HR} = \alpha(x_1) \times \Delta_{\rm x_1}
\end{equation}
where $\alpha(x_1)$ is the stretch-luminosity relation, as a function of $x_1$ \citep[e.g., see][]{Ginolin2024,Dhawan2024a} and $\Delta_{\rm x_1}$ is the difference in the $x_1$ population with redshift given by
\begin{equation}
    \Delta_{\rm x_1} = A \times \mu_1 + (1 - A) \times [a \mu_1 + (1-a) \mu_2]
\label{eq:delta_progevol}
\end{equation}
where $a = 0.47$, $\mu_1 = 0.38$ and $\mu_2 = -1.26$, i.e. the conservative case from \citet{Nicolas2021} and $A$ is given by
\begin{equation}
    A = \left(K^{-1} \times (1+z)^{2.8}  + 1 \right)^{-1}
\end{equation}
see \citet{Nicolas2021}. This model is similar to the evolution predicted by \citet{Childress2014}. We note that while there are several assumptions in the model that can be tuned differently to change the functional form of the evolution with redshift, the formalism used here is a realistic representation for testing the impact on $w_0-w_a$







\bsp	
\label{lastpage}
\end{document}

%% file: tables/bao_test.tex
\begin{table}
\caption{BAO dataset used in this analysis, which is a combination of the DESI \citet{DESI-VI} and SDSS BAO compilations.}
\centering
\begin{tabular}{|c|c|c|c|}
\hline
$D_X$ & $z_{\rm eff}$ & $\sigma$ \\
\hline 
$D_V$ & 0.106 &  0.1333 \\
$D_V$ & 0.15 &  0.168 \\
$D_M$ & 0.38 &  0.1489 \\
H$\times r_d$ & 0.38 &  280.78 \\
$D_M$ & 0.51 &  0.1827 \\
H$\times r_d$ & 0.51 &  280.78 \\
$D_M$ & 0.71 &  0.32 \\
$D_H$ & 0.71 &  0.60 \\
$D_M$ & 0.93 &  0.28 \\
$D_H$ & 0.93 &  0.35 \\
$D_M$ & 1.32 &  0.28 \\
$D_H$ & 1.32 &  0.42 \\
$D_V$ & 1.49 &  0.67 \\
$D_M$ & 2.33 &  0.75 \\
$D_H$ & 2.33 &  0.14 \\
\hline 
\end{tabular}
\label{tab:bao_data}
\end{table}

%% file: tables/sys_delt_tab.tex
\begin{table}
\caption{A summary of all the systematic groupings considered in this work and the associated amplitude of the effect as a shift in the $w_0$-$w_a$ plane. We report the effect as 1-D shifts in the $w_0$ and $w_a$ axes. }
\centering
\begin{minipage}{.5\textwidth}
    \begin{tabular}{|c|c|c|c|}
    \hline 
    Effect & Step Size (units) & $\Delta w_0$ & $\Delta w_a$ \\
    \hline 
Calib & 0.02 (mag) & 0.074 & -0.248 \\
dRV/dz & 0.5 & -0.092 & 0.234 \\
MW-Ext & Swap MW Law & -0.071 & 0.211 \\
MW-Ext2\footnote{MW-Ext is the difference between a CCM89 dust law and the F99 dust law. MW-Ext2 is between the F99 dust law and the updated, F19 dust law.} & $\ldots$ & -0.077 & 0.157 \\
Int. Scatter & Swap Scatter Model & -0.063 & 0.168 \\
IG Dust & 0.01 log $\Omega_{\rm dust}$ & -5363 $\Omega_{\rm dust}$ & 10566 $\Omega_{
\rm dust}$ \\
Prog $x_1$ & 0.1 & -0.157 & 0.292 \\
Prog $M_B$ & 0.02 (mag) & -0.058 & 0.116 \\
$\Omega_{\rm M}$ mismatch & 0.01 & -0.026 & -0.009 \\

   \hline 
    \end{tabular}
    \label{tab:sys_delt}
\end{minipage}
\end{table}